# Mid-infrared photothermal imaging flow cytometry

Yusei Sugawara[1], Keiichiro Toda[2*], Genki Ishigane[1], Masato Fukushima[1], and Takuro Ideguchi[1,2,3,4**]

[1] Department of Physics, The University of Tokyo, Tokyo, Japan

[2] Institute for Photon Science and Technology, The University of Tokyo, Tokyo, Japan

[3] Department of Advanced Materials Science, The University of Tokyo, Chiba, Japan

[4] RIKEN Center for Advanced Photonics, RIKEN, Saitama, Japan

[*] keiichiro--toda@g.ecc.u-tokyo.ac.jp

[**] ideguchi@edu.k.u-tokyo.ac.jp

**Abstract**
Imaging flow cytometry (IFC) enables high-throughput single-cell analysis but largely relies on fluorescence labeling to obtain molecular specificity. Label-free vibrational imaging can provide intrinsic chemical contrast, yet coherent Raman-based methods interrogate only a limited axial volume, which restricts quantitative whole-cell analysis under flow. Mid-infrared photothermal (MIP) microscopy offers a promising route to overcome this limitation by combining linear mid-infrared (MIR) absorption-based chemical contrast with visible-light detection, allowing chemical imaging of a broader axial volume of each cell in a wide-field configuration. However, applying MIP microscopy to rapidly flowing cells has been difficult because conventional frame-sequential acquisition of MIR-ON and MIR-OFF images is highly susceptible to motion-induced subtraction artifacts. Here we demonstrate MIP-IFC, a label-free imaging flow cytometry platform based on single-shot nanosecond-dual-pulse MIP (SNAP-MIP) microscopy. SNAP-MIP encodes the MIR-ON and MIR-OFF states into separate holographic channels within a single camera exposure, reducing their temporal separation to 20 ns. This single-shot acquisition suppresses motion artifacts and increases the allowable sample velocity for artifact-free MIP imaging by five orders of magnitude compared with conventional frame-sequential MIP imaging. Leveraging this capability, MIP-IFC acquired chemical images at 500 frames per second and achieved a cellular event rate up to ~70 events $s^{-1}$. We demonstrate quantitative chemical discrimination of flowing microbeads and apply MIP-IFC to single-cell profiling of oleic-acid-induced lipid accumulation, adipocyte differentiation, and confluence-dependent cellular heterogeneity. These results establish MIP-IFC as a high-throughput, quantitative, label-free chemical imaging platform for single-cell phenotyping under flow.

## Introduction

Imaging flow cytometry (IFC) integrates continuous flow-based cell handling with high-speed optical imaging to enable morphological and subcellular analysis across large cell populations[1-3]. By extracting hundreds of quantitative image features from each event, IFC supports high-dimensional phenotypic profiling of single cells[1,4]. Modern platforms achieve throughput of thousands of cells per second at submicrometer spatial resolution. In parallel, advances in machine-learning-based analysis have enhanced the ability to resolve subtle phenotypic differences and cellular heterogeneity[1,4-7]. Despite these advances, robust identification of cellular states and subpopulations still largely relies on fluorescence labeling. Fluorescence imaging provides molecular specificity but can introduce cytotoxic effects and perturb native cellular physiology[8, 9], limiting its applicability in contexts that require minimal sample disturbance or preservation for downstream analysis. In addition, certain cell types, such as primary immune cells, often exhibit weak or heterogeneous fluorescence signals, which further complicates reliable analysis[10,11]. These challenges motivate the development of label-free IFC modalities that can provide intrinsic contrast without external probes.

Quantitative phase imaging (QPI)-based IFC[12-15] has emerged as a powerful approach for high-throughput label-free characterization of cells under flow. By measuring optical path length contrast, QPI provides quantitative information on cell morphology, dry mass, and intracellular structural organization, enabling label-free analysis of cellular biophysical properties and phenotypic heterogeneity at throughput approaching that of fluorescence-based IFC. QPI-based IFC has developed along several technical directions, including off-axis digital holography (DH)[14] and time-stretch imaging[15]. Among these approaches, DH is particularly useful for flow imaging because it records the complex optical field in a single acquisition. The measured complex field can then be numerically propagated through free space, allowing QPI images to be computationally refocused without mechanical axial scanning. This capability reduces sensitivity to axial variations in cell position. Despite these advantages, phase imaging inherently lacks chemical specificity. QPI-based IFC alone is therefore insufficient to resolve molecular composition and chemically distinct cellular states.

Vibrational microscopy offers a complementary label-free strategy by extracting chemical contrast from endogenous molecular vibrations, primarily via Raman scattering and mid-infrared (MIR) absorption[16,17]. In the context of IFC, coherent Raman techniques have enabled high-throughput single-cell chemical analysis, including quantification of metabolic heterogeneity and label-free identification of cancer cells at throughputs of ~100 cells per second[18,19]. However, these methods rely on nonlinear optical processes that confine signal generation to a thin axial region near the focal plane, typically on the order of one to a few micrometers. As a result, they probe only a limited subcellular volume rather than the full axial extent of each cell. Although acoustic focusing and related microfluidic techniques improve cell alignment, residual axial variations in cell position can still change which subcellular region overlaps with the thin excitation volume near the focal plane. Even under ideal alignment, this axially confined volume samples only a local portion of each cell, making quantitative readouts susceptible to cell-to-cell differences in the spatial organization of organelles and chemical content. These limitations highlight the need for alternative vibrational imaging strategies that can probe the full cellular volume.

Mid-infrared photothermal (MIP) microscopy offers a promising route to address this need by combining MIR absorption-based chemical contrast with visible-light detection[20-32]. In this approach, MIR absorption induces a local refractive index (RI) change that is detected using visible-light microscopy, enabling MIR absorption imaging with visible diffraction-limited resolution. Because this contrast arises from linear MIR absorption rather than nonlinear focal-volume excitation, MIP imaging is not intrinsically confined to a thin signal-generation volume and can be implemented in a wide-field (WF) configuration. In this format, WF-MIP provides a projection-like readout of photothermal contrast over the cell thickness, allowing chemical information to be collected from a broader axial volume than a restricted focal slice. When DH is used for visible detection in this WF-MIP framework, MIP-DH additionally provides QPI-based morphological information and enables computational refocusing within a single setup. Despite these advantages, applying WF-MIP imaging, including MIP-DH, to rapidly flowing samples remains challenging. In conventional implementations, images acquired with and without MIR illumination are recorded in separate camera exposures, typically separated by >1 ms, and MIP contrast is obtained from their difference[21-32]. This sequential acquisition is inherently sensitive to sample motion and therefore introduces artifacts into reconstructed images. Consequently, avoiding motion artifacts is expected to limit the allowable flow velocity to only several tens of micrometers per second, far below the range required for high-throughput IFC.

In this work, we demonstrate MIP-DH-based IFC by suppressing motion artifacts during MIP imaging of rapidly flowing samples. This platform enables robust extraction of MIP contrast from flowing cells and supports high-throughput, label-free chemical and morphological analysis at the single-cell level. To achieve this, we integrate frequency-multiplexed off-axis DH[33] with synchronized nanosecond MIR and visible pulses. In this configuration, MIR-ON and MIR-OFF states are encoded as separate holographic channels within a single camera exposure, with an effective temporal separation of only 20 ns. The differential MIP signal is therefore extracted from near-simultaneous measurements, suppressing motion-induced inconsistencies. This approach extends the allowable flow velocity to the meter-per-second range, reaching values comparable to those of fluorescence-based IFC platforms. Using this platform, we acquired MIP images at 500 frames per second (fps) and achieved a cellular event rate of ~70 events $s^{-1}$ (eps). We then demonstrated quantitative chemical discrimination in flowing samples and applied MIP-IFC to single-cell profiling of oleic-acid-induced lipid accumulation, adipocyte differentiation, and confluence-dependent cellular heterogeneity.

## Results

### Single-shot nanosecond-dual-pulse MIP (SNAP-MIP) microscopy for IFC.

We first describe the operating principle of MIP-DH[24] and the conventional acquisition scheme used to extract the MIP signal. In MIP-DH, wide-field MIR irradiation induces localized heating at MIR-absorbing regions within the field of view, generating transient RI changes via the thermo-optic effect. Off-axis DH detects these RI changes as phase shifts. In this configuration, interference between the object beam transmitted through the sample and a tilted reference beam generates a hologram on the image sensor (Fig. 1a). Fourier transformation of this hologram separates the interferometric terms into distinct spatial frequency bands, allowing the sideband that contains the complex object-field information to be isolated. For MIP imaging, the complex fields are reconstructed from MIR-ON and MIR-OFF holograms, acquired with and without MIR excitation, respectively, and the MIP contrast is obtained as the difference between the corresponding phase images. This acquisition scheme provides MIP contrast together with a quantitative phase image from the MIR-OFF state, enabling dual-modal analysis of chemically specific contrast and cellular morphology. Because off-axis DH preserves the complex optical field, both the quantitative phase image and the MIP image can be computationally refocused without mechanical adjustment of the sample position (see "Reconstruction procedure" in Methods and in Supplementary Note 1 for details).

In conventional MIP-DH systems, MIR-ON and MIR-OFF images are recorded in separate sensor frames using successive probe pulses, which are typically separated by ~1 ms or longer (Fig. 1b). This millisecond-scale separation typically arises from the ~1-kHz repetition rate used in common MIP-DH implementations to suppress thermal accumulation from repeated MIR excitation. When the sensor frame interval exceeds the pulse-to-pulse interval, the sensor frame interval instead determines the temporal separation between the two phase images. For static samples, this temporal separation is not problematic because subtracting the MIR-OFF phase image from the MIR-ON phase image removes the background phase and isolates the MIP-induced phase change (Fig. 1b, "static"). For moving samples, however, sample displacement during the interval between the two acquisitions causes a spatial mismatch between the MIR-ON and MIR-OFF phase images. This mismatch generates subtraction artifacts in the reconstructed MIP image (Fig. 1b, "moving"). In IFC, such motion-induced artifacts severely restrict the allowable flow velocity to only several tens of μm per second, corresponding to a typical throughput of ~0.1 eps[34](see Supplementary Note 2). Reducing the temporal separation between the MIR-ON and MIR-OFF acquisitions is therefore essential. However, in frame-sequential acquisition, this interval cannot be reduced below the limits imposed by the laser repetition period and the sensor frame interval.

To overcome these limitations, we developed a method termed single-shot nanosecond-dual-pulse MIP (SNAP-MIP) microscopy. SNAP-MIP addresses the repetition-rate constraint by generating two temporally separated replicas from a single ~10-ns visible probe pulse. The first replica reaches the sample immediately before the ~10-ns MIR excitation pulse and probes the MIR-OFF state, whereas the second replica reaches the sample immediately after MIR excitation and probes the MIR-ON state. This dual-pulse design reduces the effective temporal separation between the two phase measurements to ~20 ns (Fig. 1c). SNAP-MIP also removes the frame-rate constraint of the image sensor by recording the MIR-OFF and MIR-ON states within a single sensor exposure. The two states are separated computationally using spatial-frequency-multiplexed off-axis DH[33,35]. In this scheme, two reference beams with different incidence angles encode the MIR-ON and MIR-OFF information into separate spatial-frequency bands within a single hologram (Fig. 1d). Fourier transformation of the recorded hologram separates the two encoded sidebands, which are then isolated and filtered independently. Inverse Fourier transformation of each sideband reconstructs the corresponding phase image, and the MIP image is obtained by subtracting the MIR-OFF phase image from the MIR-ON phase image. By combining nanosecond-scale dual-pulse probing with single-exposure holographic multiplexing, SNAP-MIP theoretically increases the permissible flow velocity for motion-artifact-free imaging to approximately 1 m $s^{-1}$ (see Supplementary Note 2). This capability enables MIP imaging of rapidly flowing samples (Fig. 1c, "moving").

A schematic of the SNAP-MIP-based IFC system is shown in Fig. 1e. A Q-switched Nd:YAG laser (1,064-nm wavelength, 1 kHz repetition rate, ~10-ns pulse duration) serves as the common pump source for MIR excitation and visible probe generation. The MIR excitation pulses are generated using a tunable optical parametric oscillator (OPO) operating at 2,400-3,600 $cm^{-1}$ and delivered to the sample plane with a pulse energy of approximately 5.6 μJ. The visible probe pulses are generated by second harmonic generation (SHG) at 532-nm. The visible output is then split into two pulses separated by ~20 ns, which interrogate the MIR-ON and MIR-OFF states, respectively. The two probe pulses are introduced into a spatial-frequency-multiplexed off-axis DH interferometer based on a Mach-Zehnder configuration. In the sample arm, specimens flow through a 40 µm-thick glass capillary without hydrodynamic focusing. The MIR excitation and visible probe beams illuminate an approximately 40 µm × 40 µm field of view. After passing through the sample, each probe pulse interferes with its corresponding reference beam, generating a spatial-frequency-multiplexed hologram that encodes both MIR-OFF and MIR-ON information. The hologram is recorded by an image sensor operating at 500 fps, which determines the MIP imaging frame rate. A detailed optical layout is provided in "Detailed schematic of SNAP-MIP-based IFC system" in Methods, and in Supplementary Note 3.

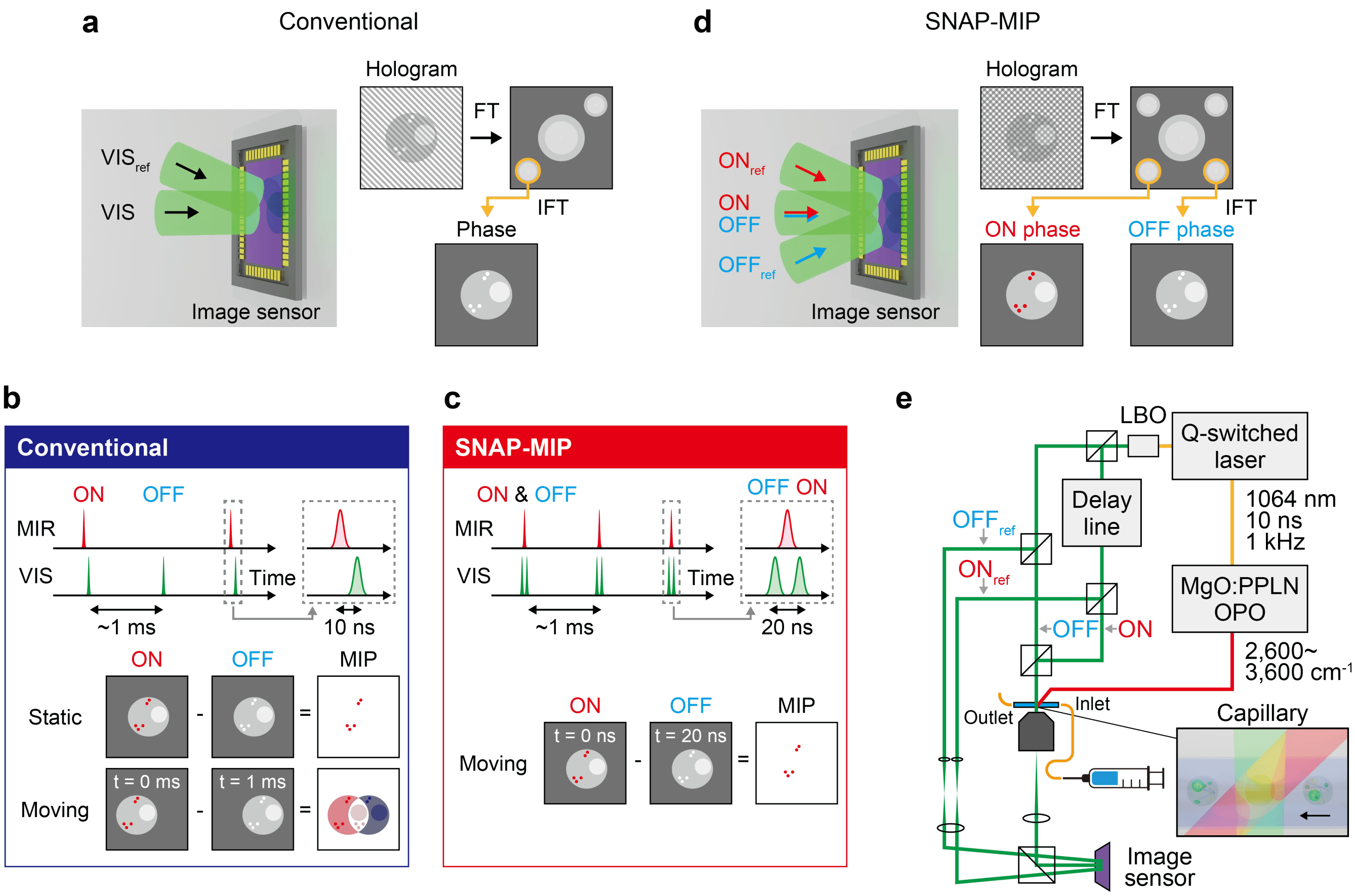

**Fig.1 Principle and schematic of SNAP-MIP microscopy for IFC. a,** Operating principle of conventional MIP-DH. FT, Fourier transform; IFT, inverse Fourier transform; **b,** Conventional frame-sequential MIP acquisition scheme (top) and MIP images for static and moving samples (bottom). **c,** SNAP-MIP acquisition scheme (top) and MIP images for moving samples (bottom). **d,** Spatial-frequency multiplexing in SNAP-MIP for single-exposure recording of the MIR-OFF and MIR-ON states. **e,** Schematic of the SNAP-MIP-based IFC system. PPLN, periodically poled lithium niobate; LBO, lithium triborate.

**Validation of motion-artifact suppression by SNAP-MIP microscopy.**

To evaluate the ability of SNAP-MIP microscopy to suppress motion-induced artifacts, we compared MIP images of moving polymethyl methacrylate (PMMA) beads acquired using conventional MIP-DH and SNAP-MIP. The MIR excitation was tuned to 2,925 cm$^{-1}$, corresponding to the $CH_2$ stretching vibration of PMMA. A MIP image of a static PMMA bead was used as the motion-artifact-free reference (Fig. 2a). When the sample stage was translated laterally at 200 µm s$^{-1}$, conventional MIP-DH produced severely distorted MIP images (Fig. 2b, top). The artifacts appeared as alternating positive and negative features along the direction of motion, consistent with relative spatial mismatch between the MIR-ON and MIR-OFF acquisitions. Under the same conditions, SNAP-MIP preserved image quality comparable to that of the static reference and accurately reproduced both the bead contrast and lateral displacement of the bead (Fig. 2b, bottom). SNAP-MIP also remained robust at substantially higher sample velocities. At approximately 8 cm $s^{-1}$, comparable to the speed used in subsequent IFC experiments, SNAP-MIP still produced clear MIP images without apparent motion-induced distortion (see Supplementary Note 4). These results demonstrate that SNAP-MIP suppresses motion artifacts and enables faithful MIP imaging of moving samples.

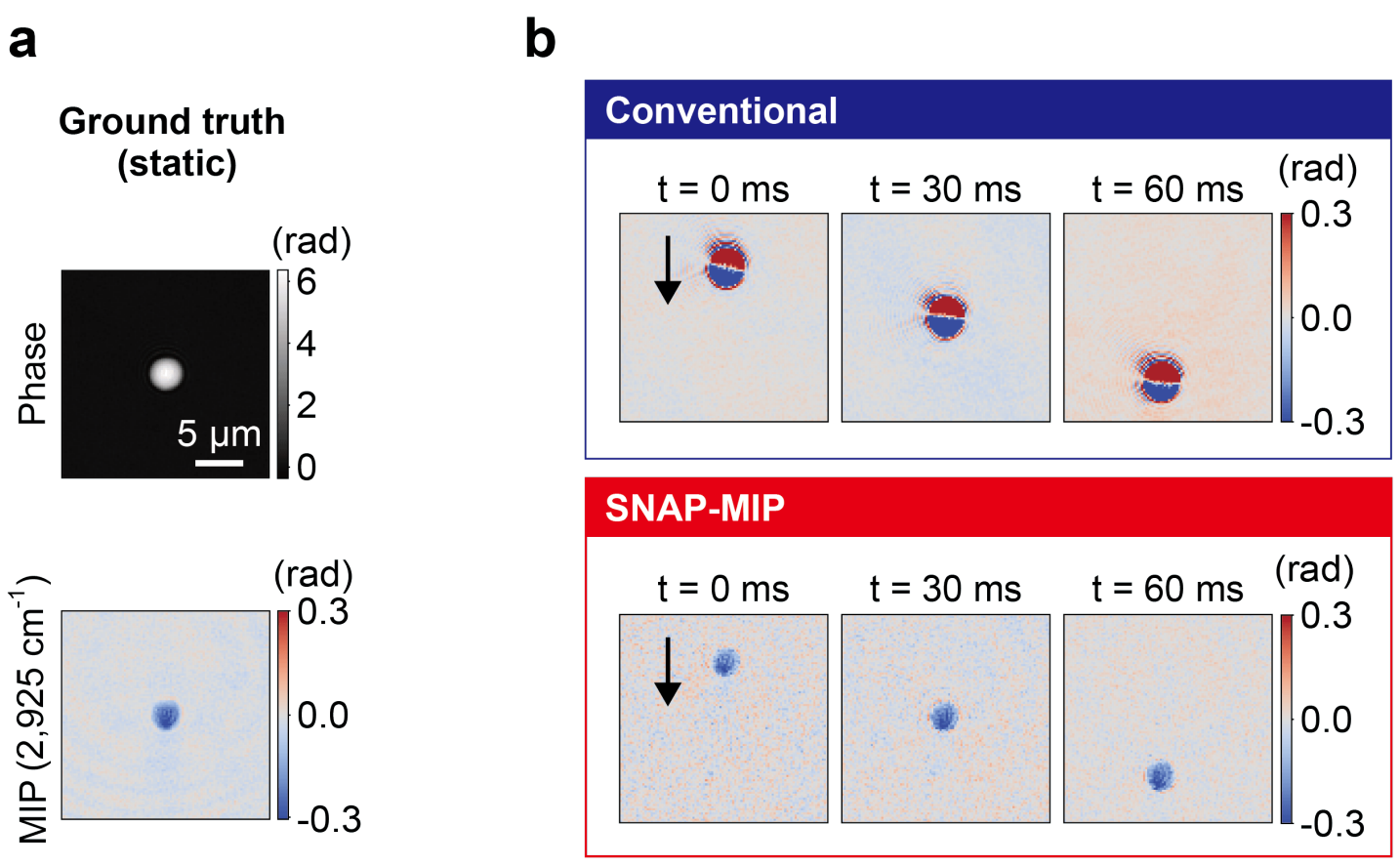

**Fig. 2 Validation of motion-artifact suppression by SNAP-MIP microscopy. a,** Motion-artifact-free reference images of a static PMMA bead. Top, phase image. Bottom, MIP image at 2,925 $cm^{-1}$. **b,** Comparison of MIP imaging of a laterally moving PMMA bead acquired by conventional MIP-DH (top) and SNAP-MIP (bottom).

**Quantitative chemical discrimination of flowing microbeads.**

We next evaluated whether SNAP-MIP enables quantitative chemical discrimination under flow. A mixture of PMMA and silica beads was flowed through a glass capillary and imaged by SNAP-MIP. The MIR excitation was tuned to 2,925 $cm^{-1}$, corresponding to the $CH_2$ stretching vibration of PMMA. The flow rate was set to 42 cm $s^{-1}$, yielding a bead event rate of 179 eps (see "Flow procedure" and "Event-rate calculation" in Methods and Supplementary Note 5). Representative MIR-OFF phase images and the corresponding MIP images of the two bead types are shown in Fig. 3a. A pronounced MIP signal was detected only from the resonant PMMA beads, whereas the off-resonant silica beads showed no clear corresponding response. This bead-type-dependent response confirms chemically selective MIP imaging under flow. Although silica beads did not produce a clear photothermal signal, weak residual edge-like features were occasionally observed in their reconstructed MIP images. Control measurements indicated that these residual features were not caused by MIR absorption or sample motion, but were artifacts originating from the spatial-frequency-multiplexed off-axis DH scheme (see Supplementary Note 6 for details). We therefore refer to these residual features hereafter as multiplex artifacts.

We then examined whether computational refocusing preserves accurate MIP reconstruction for beads flowing at different axial positions. Figure 3b shows representative phase and MIP images of beads located at different axial positions before and after computational refocusing. Before refocusing, axial defocus degraded both the phase and MIP images, particularly for beads displaced from the focal plane. Computational refocusing restored clear bead profiles and yielded consistent QPI and MIP contrast across different axial positions. These results show that SNAP-MIP can recover reliable MIP contrast over an extended axial range from a single recorded hologram, without mechanical axial alignment. This relaxes the requirement for precise axial confinement by acoustic focusing or related microfluidic techniques, enabling robust chemical imaging of samples passing through different axial positions within the capillary.

Figure 3c quantitatively characterizes the bead populations using histograms and scatter plots of phase values and MIP signal intensities. The phase values were obtained from computationally refocused QPI images, whereas MIP signal intensities were obtained from the computationally refocused MIP images after normalization by the measured MIR fluence distribution, which allowed MIP contrast to be evaluated independently of spatial variations in MIR excitation fluence (see "Calibration and normalization of MIR fluence distribution" in Methods). The measured phase values were 5.91 ± 0.24 rad for PMMA beads and 3.61 ± 0.08 rad for silica beads. These values agree well with theoretical estimates based on the refractive indices of the beads (1.49 for PMMA and 1.43 for silica) and their nominal diameters (3.04 μm for PMMA and 2.86 μm for silica). The variation in phase values was primarily attributed to the intrinsic size dispersion of the commercial beads (3.04 ± 0.11 μm for PMMA and 2.86 ± 0.06 μm for silica). These results indicate that the two bead populations can be reliably distinguished using phase information. Consistent with this phase-based classification, the MIP signal intensities provided clear chemical separation between PMMA and silica beads, with values of -0.52 ± 0.06 rad/(J cm$^{-2}$) and -0.01 ± 0.06 rad/(J cm$^{-2}$), respectively. The variance in the MIP signal was dominated by phase noise, which arose mainly from optical shot noise and multiplex artifacts (see Supplementary Note 7 for details). These results establish that SNAP-MIP enables quantitative chemical discrimination of flowing samples with distinct molecular compositions.

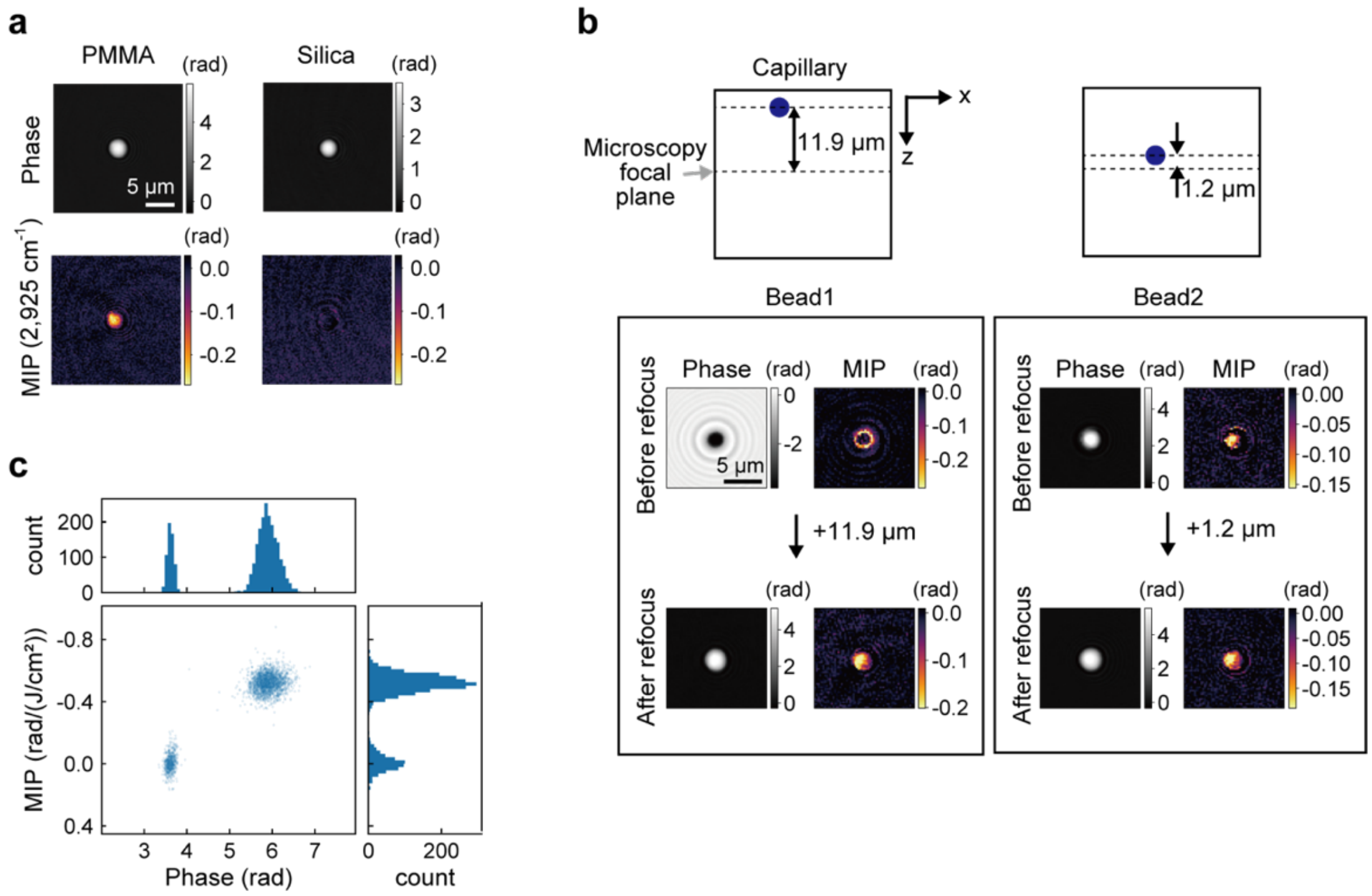

**Fig. 3 Quantitative chemical discrimination of flowing microbeads by SNAP-MIP. a,** Representative phase images and corresponding MIP images at 2,925 cm$^{-1}$ of flowing PMMA and silica beads. **b,** Computational refocusing of phase and MIP images for PMMA beads flowing at different axial positions. The y-direction corresponds to the flow direction, whereas the z-direction corresponds to the propagation direction of the visible probe beam. **c,** Quantitative classification of PMMA and silica beads based on phase values and MIR-fluence-normalized MIP signal intensities. Histograms show the marginal distributions of each feature, calculated from 2,512 beads.

**MIP-IFC analysis of lipid accumulation under flow.**

We next examined whether MIP-IFC can discriminate cell populations with distinct chemical compositions under flow. To generate controlled compositional differences, we induced intracellular lipid accumulation in COS-7 cells by oleic acid treatment under three conditions, 0, 50, and 100 μL (see "Cell preparation of COS-7 cells and oleic acid treatment" in Methods for details)[36]. The MIR excitation wavenumber was set to 2,925 cm$^{-1}$, corresponding to the $CH_2$ stretching absorption band, a lipid-associated vibrational signature. The flow rate was set to 10 cm s$^{-1}$, yielding a cellular event rate of 70, 42 and 46 eps, with n = 12,627, 7,515 and 8,310 cells for the 0, 50 and 100 μL conditions, respectively (see Supplementary Note 5). Representative phase and MIP images for each condition are shown in Fig. 4a. The phase images mainly reflect the depth-integrated dry mass density, with contributions from proteins, lipids, and nucleic acids, whereas the MIP images preferentially highlight lipid-rich intracellular regions. Consistent with these different contrast origins, the phase images showed no obvious differences across the three conditions, whereas the MIP images of oleic acid-treated cells exhibited contrast concentrated in discrete intracellular regions, in line with localized lipid accumulation[37].

To quantify these trends, we extracted multiple features from the phase and MIP images and compared them across conditions (see "Calculation of cell feature values" in Methods for details). From the phase images, we quantified the cell diameter and dry mass density for each cell. By combining the MIP images with cell-size information, we further calculated the MIP signal density, defined as the mean RI change per cell. We also calculated the normalized Shannon entropy of the pixel-wise MIP signal-distribution for each MIP image to assess whether the intracellular MIP signal intensity was concentrated in a limited subset of pixels or distributed more uniformly across pixels (hereafter referred to as entropy). In Fig. 4b, the left column shows violin plots of the feature distributions under each treatment condition, while the right column shows the corresponding mean standard deviation (s.d.) values. Cell diameter remained largely unchanged across the three oleic acid conditions, whereas dry mass density increased only modestly in the 50 and 100 μL conditions, reaching 1.03-fold and 1.16-fold relative to the 0 μL condition, respectively. In contrast, MIP signal density increased markedly, reaching 3.87-fold and 6.80-fold higher than the 0 μL condition in the 50 and 100 μL conditions, respectively. This difference between phase-derived dry mass density and MIP signal density indicates that the MIP signal density more clearly captures lipid-associated compositional changes than dry mass density, which integrates contributions from multiple biomolecular components.

Entropy analysis further showed that entropy was lower in both the 50 and 100 μL conditions than in the 0 μL condition, indicating that the intracellular lipid-associated signal became more concentrated in a smaller subset of pixels after oleic acid treatment. This trend is consistent with the accumulation of excess intracellular lipids in lipid droplet-like storage structures. However, entropy did not differ substantially between the 50 and 100 μL conditions. Interpreted together with the increase in MIP signal density, this result suggests that higher oleic acid treatment increased the overall lipid accumulation while the heterogeneity of the MIP signal intensity across pixels remained similar between the two treated conditions. These results show that MIP-IFC can visualize and quantify differences in cellular chemical composition under flow, particularly differences in lipid content across the cell population.

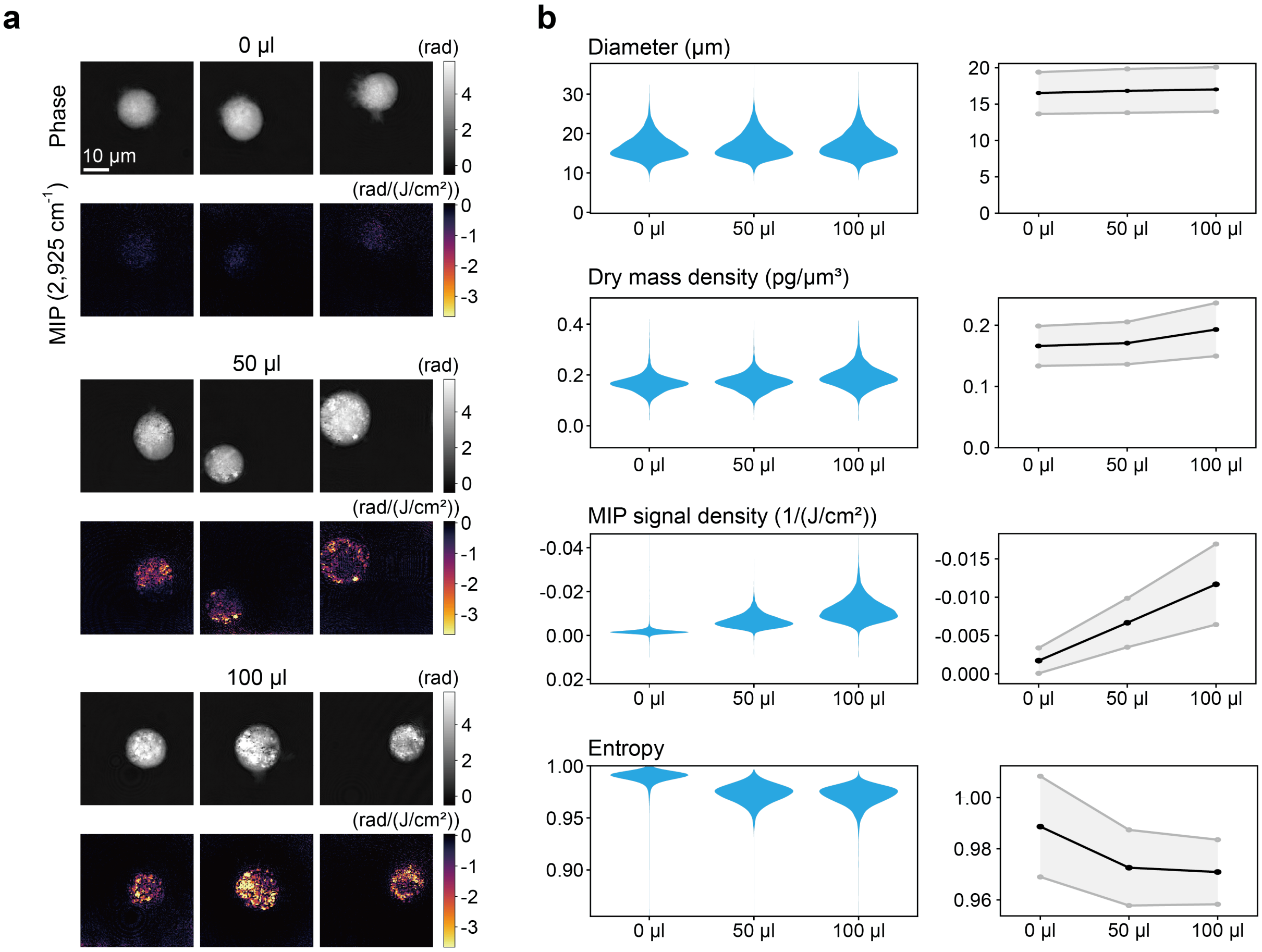

**Fig. 4 MIP-IFC analysis of oleic-acid-induced lipid accumulation under flow. a,** Representative phase and MIP images of flowing COS-7 cells treated with 0, 50 or 100 µL oleic acid. MIP images were acquired at 2,925 cm$^{-1}$. **b,** Quantitative feature analysis of COS-7 cells under each treatment condition. Left, violin plots showing the distributions of cell diameter, dry mass density, MIP signal density and entropy. Right, corresponding mean ± s.d. values.

**Single-cell analysis of adipocyte differentiation by MIP-IFC.**

We next examined whether MIP-IFC can visualize lipid accumulation associated with adipocyte differentiation at the single-cell level. To generate cells with increased intracellular lipid content, we induced adipocyte differentiation in 3T3-L1 cells, a well-established adipogenesis model that accumulates lipids after differentiation induction (see "Differentiation of 3T3-L1 cells" in Methods for the detailed procedure)[38]. Measurements were performed at a cellular event rate of 6.3 eps, and 2,270 cells were analyzed. Using the phase images together with MIP images acquired at 2925 cm$^{-1}$, we quantified the cell area and lipid-rich area for individual cells. Cell area was calculated from the phase images. Lipid-rich area was defined as the total area of intracellular structures with MIP signals above a threshold determined from the undifferentiated cell population. Because the post-induction population still contained undifferentiated cells, cells without detectable lipid-rich areas were excluded from the subsequent differentiated-cell analysis (see "Exclusion of undifferentiated-like 3T3-L1 cells" in Methods for the details).

Figure 5a shows the relationship between cell area and lipid-rich area for individual cells. The reference lines indicate lipid area fractions of 20%, 40%, and 70% relative to the cell area. The differentiated-cell population was broadly distributed along both axes, indicating substantial cell-to-cell heterogeneity in both cell size and lipid accumulation. This distribution shows that the differentiation state of 3T3-L1 cells cannot be adequately described by cell size or lipid amount alone, but requires joint evaluation of cell area, lipid-rich area, and lipid area fraction.

Figure 5b shows representative phase and MIP images of cells selected from the regions indicated in Fig. 5a, labeled S1 to S3 and L1 to L3. The representative MIP images provide a visual counterpart to the heterogeneity observed in Fig. 5a. Among cells with similar cell areas, the MIP images revealed substantial differences in the size and extent of lipid-rich structures, demonstrating that MIP-IFC can directly visualize heterogeneity in differentiation-associated lipid accumulation within the population. Comparison across the selected regions further showed that, in both the smaller-cell group (S1 to S3) and the larger-cell group (L1 to L3), the increase in lipid-rich area from region 1 to region 3 was not characterized by a uniform increase in the number of lipid structures of similar size. Instead, lipid accumulation was associated with the emergence and growth of larger lipid droplet-like structures. These observations indicate that MIP-IFC captures not only increases in total lipid content but also changes in the spatial organization of lipid storage.

In Fig. 5c, the data are replotted with cell area on the x axis and lipid area fraction on the y axis. The upper boundary of the lipid area fraction was not uniform across cell sizes, as indicated by the blue dashed line. It decreased with increasing cell area, from approximately 90% at 200 $\mu m^2$ to approximately 60% at 500 $\mu m^2$. This size-dependent decrease suggests that, within the heterogeneous differentiated cell population, the upper limit of lipid accumulation does not simply scale with cell area. This interpretation is discussed further in the Discussion section. Overall, these findings show that MIP-IFC enables single-cell quantification of lipid accumulation during adipocyte differentiation and reveals heterogeneity in differentiation state that cannot be captured by cell size alone or lipid-rich area alone.

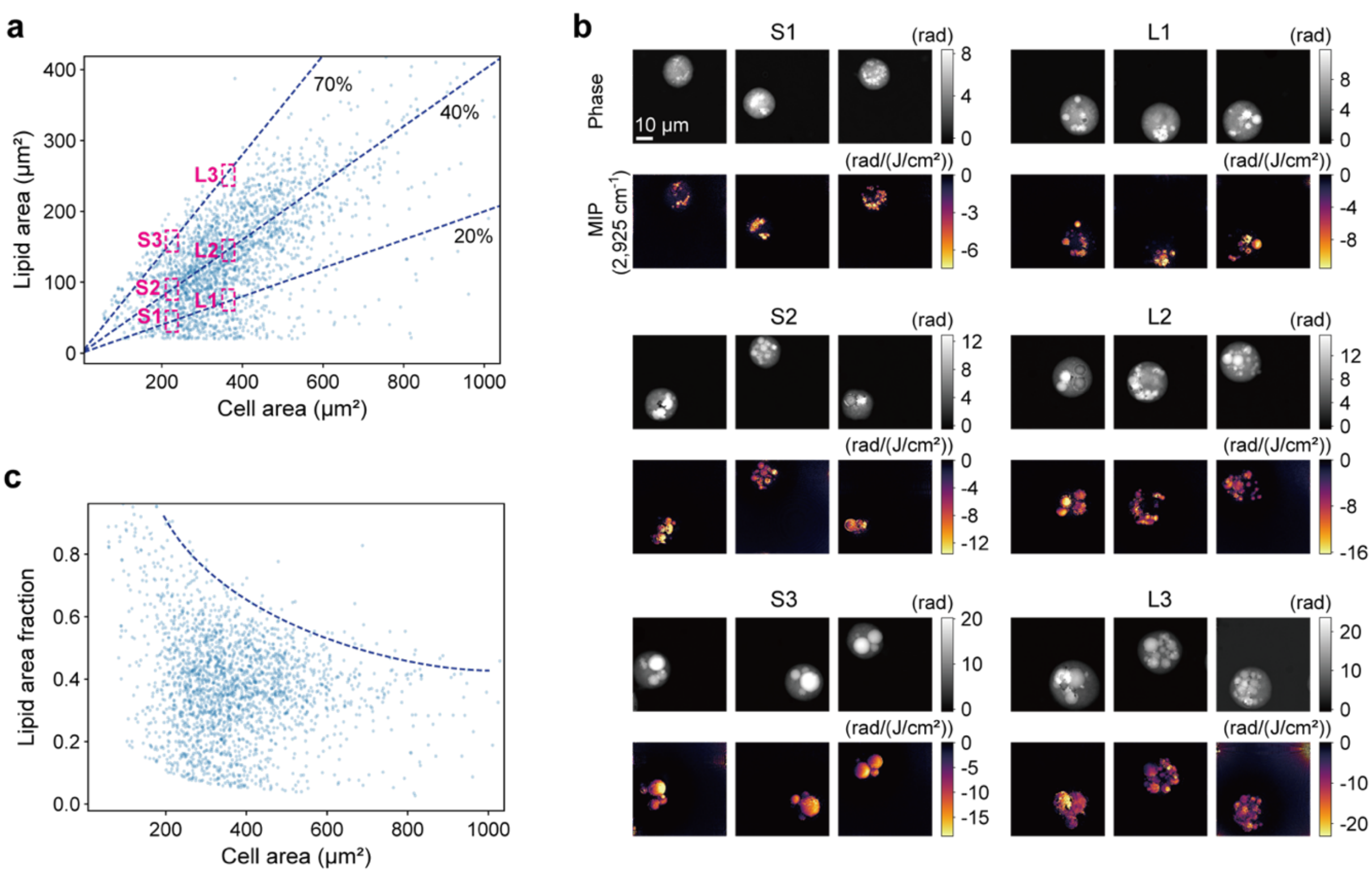

**Fig. 5 Single-cell analysis of adipocyte differentiation by MIP-IFC. a,** Relationship between cell area and lipid-rich area for differentiated 3T3-L1 cells. Dashed lines indicate lipid area fractions of 20%, 40% and 70%. **b,** Representative phase and MIP images of cells selected from the regions indicated in **a**. Cells are grouped by smaller-cell regions (S1 to S3) and larger-cell regions (L1 to L3). **c,** Relationship between cell area and lipid area fraction. The blue dashed line indicates the apparent upper envelope of the lipid area fraction.

**Culture confluence-dependent cellular profiling by MIP-IFC.**

We next used MIP-IFC to characterize changes in cellular population state associated with increasing culture confluence. COS-7 cells were seeded in culture dishes and maintained for several days. As the cells proliferated, culture confluence gradually increased and reached nearly 100% between days 8 and 10. Cells were measured at multiple time points during this process, and changes in morphology, dry mass density, and lipid-related MIP contrast were analyzed. Representative phase and MIP images at each time point are shown in Fig. 6a, and Fig. 6b summarizes the temporal evolution of the extracted features using the same analysis as in Fig. 4.

The mean cell diameter and dry mass density remained largely unchanged throughout the experiment, indicating that the overall cell size and bulk dry-mass content were relatively stable. In contrast, MIP signal density increased from day 8 onward, indicating an increase in lipid-associated molecular content. Entropy calculated from the MIP images decreased over the same period, suggesting that the lipid-associated MIP signal became increasingly concentrated in a limited subset of pixels. This trend is consistent with the accumulation of intracellular lipids in lipid droplet-like storage structures.

The temporal changes in standard deviation further revealed an increase in population heterogeneity. After the culture reached approximately 100% confluence, the distribution of MIP signal density became progressively broader. This broadening indicates that, under high-density culture conditions, cells did not converge to a single common state but instead became increasingly heterogeneous. The possible biological origins of the observed changes in both mean values and population heterogeneity are discussed in the Discussion section. These findings show that MIP-IFC can capture both shifts in population-averaged cellular composition and the emergence of population heterogeneity under high-density culture conditions through integrated single-cell analysis of morphology, dry mass-related properties, and chemical contrast.

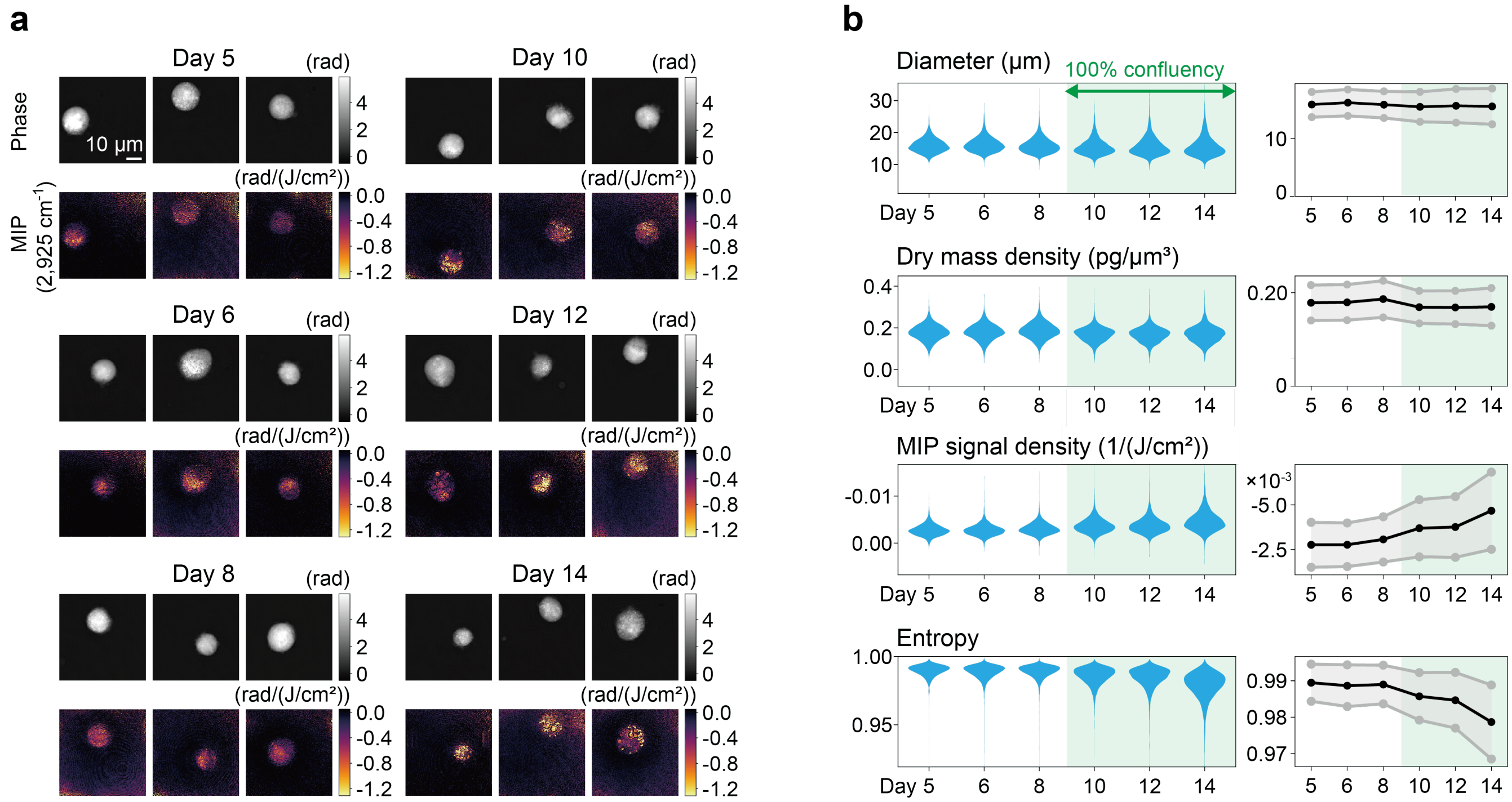

**Fig. 6 Culture-confluence-dependent cellular profiling by MIP-IFC. a,** Representative phase and MIP images of flowing COS-7 cells measured at different culture times as confluence increased. MIP images were acquired at 2,925 $cm^{-1}$. **b,** Quantitative feature analysis across culture time. Left, violin plots showing the distributions of cell diameter, dry mass density, MIP signal density and entropy. Right, corresponding mean ± s.d. values.

## Discussion

The results in Figs. 5 and 6 provide insight into heterogeneous cellular responses in two biological contexts: adipocyte differentiation and high-density culture. In Fig. 5c, the upper limit of lipid area fraction decreased with increasing cell area. Because this fraction is normalized by cell area, larger cells must contain a greater absolute lipid-rich area to reach the same fractional value as smaller cells. The reduced upper bound observed in larger cells therefore suggests that, at the time of measurement, even the most lipid-rich cells in this size range had not accumulated sufficient lipid-rich regions to reach the fractional values observed in smaller cells. This relationship indicates heterogeneous progression of lipid remodeling across cells of different sizes during adipocyte differentiation.

Figure 6 further suggests that cellular states changed during the transition to high-density culture. Near 100% confluence, several factors that can alter cellular physiology may arise simultaneously, including contact-dependent growth inhibition, nutrient depletion, accumulation of metabolic by-products, and acidification of the medium. Under these high-density culture conditions, the lipid-associated MIP signal increased, and its intensity became increasingly concentrated in a limited subset of pixels. These features are consistent with the possibility that a subset of cells shifted from a proliferation-dominant state toward a metabolically altered state associated with adaptation to high-density culture. This interpretation is consistent with reports that lipid accumulation can accompany stress responses or metabolic adaptation[39]. It is also notable that heterogeneity in lipid content increased around the time when the culture reached nearly 100% confluence. One possible explanation is that local microenvironments, such as nutrient availability, waste accumulation, and cell-cell contact, became spatially heterogeneous in dense cultures, leading to divergent cellular responses within the population.

We next discuss several technical extensions that could further enhance the performance and application scope of SNAP-MIP-based IFC. A key direction is increasing image-acquisition throughput. In the current implementation, the throughput is primarily limited by the image sensor, which operates at 500 fps for a reconstructed field of view of 279 × 279 pixels. This corresponds to a pixel throughput of $3.9\times10^{7}$ pixels $s^{-1}$, already several times higher than that of reported coherent Raman IFC systems, which typically operate at $2.0\times10^{7}$ pixels $s^{-1}$ [18]. With faster sensors, the dominant constraint would shift to the repetition rate of the MIR and visible pulse trains, currently 1 kHz. Importantly, this repetition-rate constraint is not fundamental. In conventional MIP-DH imaging of static samples, increasing the repetition rate leads to cumulative heating, which can induce thermal damage and thereby limits the allowable excitation rate. In SNAP-MIP-based IFC, by contrast, the MIR-ON and MIR-OFF states are acquired within a single excitation event, and sample flow ensures that each sample volume experiences only a single MIR excitation. This prevents cumulative heating of the same sample volume and relaxes the repetition-rate constraint. When combined with a high-speed image sensor, increasing the repetition rate by an order of magnitude could enable imaging rates approaching 10,000 fps, provided that sufficient MIR pulse energy is maintained. This strategy provides a scalable pathway toward throughput improvements exceeding one order of magnitude relative to coherent Raman IFC.

A second direction is expanding the spectral dimensionality, which is important for biochemical characterization and diagnostic applications. Within the SNAP-MIP framework, multispectral acquisition could be implemented by extending the spatial-frequency multiplexing strategy. In the current implementation, only two states, MIR-ON and MIR-OFF, are encoded into distinct spatial-frequency bands within a single hologram, leaving a substantial portion of the spatial-frequency domain unused. The unoccupied frequency bands could be used to encode additional MIR-ON states at different excitation wavelengths, enabling simultaneous acquisition of multiple spectral channels within a single exposure. This capability allows parallel analysis of multiple chemical targets under flow conditions.

Finally, we comment on the broader biological scope of MIP-IFC. The present results demonstrate that our MIP-IFC can capture integrated changes in cellular state under flow, including lipid accumulation, lipid spatial organization, morphology and dry mass-related properties. This capability supports applications in the analysis of differentiation processes, culture-environment-induced state transitions and metabolically distinct subpopulations. Further improvements in throughput and spectral dimensionality discussed above would extend this scope to more robust analysis of rare or low-abundance subpopulations and more detailed biochemical profiling. With these advances, MIP-IFC could become a platform for high-throughput, label-free chemical phenotyping in diagnostic screening and functional cell-state analysis.

## Methods

### Detailed schematic of SNAP-MIP-based IFC system.

Figure S2 shows a detailed schematic of the SNAP-MIP system. A Q-switched Nd:YAG laser (NL202, Ekspla) served as a common pump source for MIR generation and visible probe generation. Frequency doubling of the 1,064-nm output to 532 nm was performed using a 15-mm-long lithium triborate (LBO) crystal, whereas tunable MIR light spanning 2,400 to 3,600 $cm^{-1}$ was generated using a 50-mm-long fan-out periodically poled lithium niobate (PPLN) crystal (HC Photonics)[28]. The 532-nm output was split into two visible probe pulses with a temporal separation of 20 ns using a free-space delay line. The relative timing between the MIR excitation pulse and the two visible probe pulses was adjusted with an additional delay stage to ensure sequential interrogation of the MIR-OFF and MIR-ON states.

The two probe pulses were directed into a Mach-Zehnder interferometer configured for spatial-frequency-multiplexed off-axis DH. In the sample arm, the probe pulses were recombined and coupled into a short single-mode fiber (150 mm; 460HP, Thorlabs) before illuminating the specimen. This configuration ensured identical illumination wavefronts for both probes and reduced beam-pointing fluctuations associated with the long free-space delay path. The probe light transmitted through the sample was imaged onto the image sensor (VLXT-17M.I, Baumer) using a water-immersion objective (UPLSAPO60XW) and relay lenses, resulting in a total magnification of 150. The spatial resolution was 222 nm according to the Nyquist-Shannon sampling criterion. In the reference arm, two reference beams corresponding to the two probe pulses were delivered through separate fibers and directed to the sensor with different incident angles. These different incidence angles encoded the MIR-OFF and MIR-ON sample fields into separate spatial-frequency bands within a single hologram.

### Reconstruction procedure.

The reconstruction workflow is illustrated in Fig. S1. A recorded spatial-frequency-multiplexed hologram contains two separated sidebands in the Fourier domain, corresponding to the MIR-OFF and MIR-ON probe pulses. These sidebands were isolated independently and transformed back to the image domain to reconstruct the complex optical fields for the MIR-OFF and MIR-ON states.

The reconstructed fields include the sample-dependent optical phase together with sample-independent background phase offsets. Because the MIR-OFF and MIR-ON probe pulses interfere with different reference fields, the two holographic channels contain different static phase offsets. These offsets mainly originate from optical-path differences between the two reference-beam paths and can vary spatially even in the absence of a specimen. To correct for these channel-dependent offsets, we acquired a sample-free reference measurement under the same optical conditions, with only the aqueous medium present. This reference measurement was used to remove the reference-beam-dependent phase offsets from the reconstructed MIR-OFF and MIR-ON fields before computational refocusing.

The corrected complex fields were then computationally refocused using angular-spectrum propagation. The optimal focal plane was determined for bead and cell samples using the Gini index (GI) criterion[40]. After computational refocusing, phase images were obtained for the MIR-OFF and MIR-ON states. The final MIP contrast was calculated as the difference between the MIR-ON and MIR-OFF phase images.

**Flow procedure.**
Glass capillaries with thicknesses of 20 µm and 40 µm (VITROCOM) were used as flow channels for bead and cell experiments, respectively. Each capillary was mechanically stabilized by clamping it between a glass coverslip and a $CaF_2$ coverslip (see Supplementary Note 8 for details). To minimize the strong MIR absorption of $H_2O$ in both the flow medium and the immersion layer of the objective, $D_2O$ was used as the solvent. Although prolonged exposure to $D_2O$ could perturb cellular physiology, the residence time of cells within the channel during imaging was on the order of seconds. Under these conditions, no measurable impact on cellular state was expected.

For bead measurements, an appropriate amount of bead suspension was added to $D_2O$. The resulting mixture was loaded into a syringe, mounted on a syringe pump, and infused through the capillary at the desired flow rate. For cell measurements, adherent cells were detached from the culture dish by trypsinization and collected by centrifugation. After removal of the supernatant, the cell pellet was resuspended in $D_2O$-based PBS. The cell concentration was then measured, and the suspension was further diluted with $D_2O$-based PBS as needed to obtain the desired concentration. The prepared cell suspension was loaded into a syringe, mounted on a syringe pump and infused through the capillary at the desired flow rate.

**Event-rate calculation.**
The acquired data were first screened using the phase images to exclude frames in which no sample was detected. The remaining sample detections were then filtered based on phase and MIP signal to exclude invalid measurements, including debris and measurements affected by autofocus failure. For each detected sample, the local MIR fluence at its position was then estimated. Detections were excluded when this local fluence was below 50% of the maximum fluence for bead measurements or 30% for cell measurements, because such locally low-fluence measurements had poor signal-to-noise ratio. The remaining valid detections were defined as events, and the event rate was calculated as the total number of events divided by the total acquisition time for each measurement.

**Calibration and normalization of MIR fluence distribution.**
To correct for fluence-dependent variations in the MIP signal, calibration data were acquired before the main bead or cell measurements to estimate the three-dimensional (3D) MIR fluence distribution. For this calibration, PMMA beads were measured under flow. From the acquired image set, we extracted the lateral and axial positions of each bead together with its MIP signal intensity. These paired position-intensity data were used as training data for Gaussian process regression, which estimated a continuous 3D function describing the position-dependent MIP response.

Because PMMA beads of the same material and nominal size produce MIP signals proportional to the local MIR fluence, this fitted function represents the relative 3D MIR fluence distribution. Using the independently measured MIR pulse energy, this relative distribution was converted into absolute fluence maps. The MIP signals obtained in the main measurements were then normalized using the corresponding fluence maps to correct for fluence-dependent signal variations. This calibration procedure and the resulting fluence maps are summarized in Fig. S8 in Supplementary Note 9.

**Cell preparation of COS-7 cells and oleic acid treatment**.
COS-7 cells were cultured in a 90-mm cell-culture dish. Cells were maintained in high-glucose Dulbecco's modified Eagle medium (DMEM; FUJIFILM Wako) containing L-glutamine, phenol red, and HEPES, supplemented with 10% fetal bovine serum (FBS; Cosmo Bio) and 1% penicillin-streptomycin-L-glutamine solution (FUJIFILM Wako). Cell cultures were maintained at 37 °C in a humidified incubator with 5% $CO_2$. For oleic acid treatment, 0, 50, or 100 µL of 4 mmol $L^{-1}$ oleic acid solution was added to the culture medium one day before measurement, and the cells were incubated overnight.

**Calculation of cell feature values.**

For cell-size estimation, a cell mask was first generated from the phase image. The phase image was normalized using percentile-based intensity scaling, followed by Gaussian blurring and Otsu thresholding to segment the cell region. The largest connected component was selected as the cell mask and refined by morphological closing and opening. The cell diameter was calculated by extracting the outer contour of the mask, measuring the enclosed area, and determining the diameter of the area-equivalent circle.

For dry-mass-density analysis, the phase values within the masked cell region were summed and converted to total dry mass using the following equation[41]:

$$M = \frac{\lambda}{2\pi\alpha} \sum_{(i,j)\in A} \phi_{ij} \Delta x \Delta y$$

where $M$ is the total dry mass, $A$ denotes the masked cell region, $\phi_{ij}$ is the phase value at pixel $(i, j)$, $\Delta x$ and $\Delta y$ are the lateral pixel sizes, $\lambda = 532$ nm is the probe wavelength, and $\alpha = 0.18$ mL $g^{-1}$ is a specific refractive increment. The cell volume was then estimated from the effective diameter calculated above, assuming a spherical cell geometry. Dry mass density was calculated as the total dry mass divided by this estimated volume.

To calculate the MIP signal density, the MIP phase values within the masked cell region were area-integrated as

$$S_{MIP} = \sum_{(i,j)\in A} \Delta\phi_{MIR,ij} \Delta x \Delta y,$$

where $\Delta\phi_{MIR,ij}$ is the MIP-induced phase change at pixel $(i, j)$. Because each MIP phase value represents an axial integral of the MIR-induced optical path-length change, this area-integrated signal is proportional to the volume integral of the RI change over the cell. This integrated signal was divided by the estimated cell volume to obtain the mean RI change, which we define as the MIP signal density.

Normalized Shannon entropy of the pixel-wise MIP signal-distribution was calculated to assess whether the intracellular MIP signal intensity was concentrated in a limited subset of pixels or distributed more uniformly across pixels. For each pixel in the masked cell region, the absolute MIP signal was divided by the sum of absolute MIP signals over the entire cell region to obtain a normalized non-negative weight,

$$p_i = \frac{|\Delta\phi_{MIR,ij}|}{\sum_{i\in A}|\Delta\phi_{MIR,ij}|}.$$

Normalized entropy was defined as

$$H = \frac{-\sum_{i\in A} p_i \log p_i}{log\ N},$$

where $A$ denotes the masked cell region and $N$ is the number of pixels in $A$.

**Differentiation of 3T3-L1 cells.**

3T3-L1 cells were seeded in culture dishes at a density of $3 \times 10^5$ cells per dish in 12 mL of growth medium. The growth medium consisted of low-glucose DMEM (Wako 041-29775) supplemented with calf serum (Sigma-Aldrich, C8056-100ML) and antibiotics. Cells were cultured for approximately 4 days until they reached 100% confluence.

For adipocyte differentiation, the growth medium was then replaced with induction medium. This induction medium was prepared from high-glucose basal medium (Wako 048-30275) supplemented with FBS, antibiotics, 0.5 mM 3-isobutyl-1-methylxanthine (I5879, Sigma-Aldrich), 0.25 μM dexamethasone (D4902, Sigma-Aldrich) and 10 μg $mL^{-1}$ insulin (093-06471, Wako). Cells were cultured in this induction medium for 2 days. Thereafter, the medium was replaced every 2 days with maintenance medium prepared from the same high-glucose basal medium supplemented with FBS, antibiotics, and 1 μg $mL^{-1}$ insulin. Measurements were performed 9 days after the initiation of differentiation.

**Exclusion of undifferentiated-like 3T3-L1 cells.**
To exclude undifferentiated-like cells from the post-induction 3T3-L1 population, we used a two-step thresholding procedure based on thresholds defined from the pre-induction dataset. First, a pixel-level MIP-signal threshold was used to generate lipid-rich area masks. Pixels with MIP signal values above this threshold were classified as lipid-rich regions, where the threshold was set so that the lipid-rich area in most pre-induction cells was zero or nearly zero. However, some pre-induction cells still contained small above-threshold regions, reflecting basal lipid-associated MIP signals. Therefore, the pixel-level threshold alone was insufficient to distinguish post-induction cells with differentiation-associated lipid accumulation from undifferentiated-like cells retaining only basal lipid signals.

We therefore defined a second threshold at the cell level based on lipid-rich area. Using the pixel-level MIP-signal threshold defined above, we calculated the lipid-rich area for each cell in the pre-induction dataset. The resulting per-cell lipid-rich area distribution was used to define a cell-level threshold that excludes cells whose lipid-rich area remains within the range observed before differentiation induction. The same pixel-level and cell-level thresholds were then applied to the post-induction dataset, and only cells with lipid-rich areas above the cell-level threshold were included in the differentiated-cell analysis.

**Data, Materials, and Software Availability**
The data provided in the manuscript is available from the corresponding author upon reasonable request.

**Acknowledgments**
This work was financially supported by Japan Society for the Promotion of Science (23H00273, 25H01386, T.I.), JST FOREST Program (JPMJFR236C, T.I.), RIKEN TRIP initiative (T.I.). Y. S. was supported by the FoPM WINGS Program of The University of Tokyo and JST SPRING (JPMJSP2108).

**Author contributions**
T.I. conceived the project. K.T. designed the SNAP-MIP methodology. Y.S. performed the experiments and analyzed the data. Y.S. established the cell-flow procedure. Y.S., K.T., G.I., M. F., and T.I. discussed the implementation of the results. K.T. and T.I. supervised the work. Y.S., K.T., and T.I. wrote the manuscript with inputs from other authors.

**Competing interests**
Y.S., K.T., M. F., and T.I. are inventors of a filed patent application related to the SNAP-MIP methodology (optical microscopy, JP 2025-35847). This application covers the methodology reported here. G.I. declares no competing interests.